\documentclass{article}       
\usepackage{e}                

\usepackage{graphicx}
\usepackage{amsmath}

\begin{document}
\branch{C}   
%
\title{Heavy Ions at LHC: Theoretical Issues}
\author{R.~J.~Fries\inst{1} \and B.~M\"uller\inst{1}}
\institute{Department of Physics, Duke University, Durham, NC 27708 \\
 e-mail: rjfries@phy.duke.edu}
\maketitle
\begin{abstract}
We give a brief overview of our current theoretical understanding of 
ultra-relativistic heavy ion collision and the properties of super-hot nuclear
matter. We focus on several issues that have been discussed in connection with
experimental results from the CERN SPS and from the Relativistic Heavy Ion
Collider RHIC. We give an extrapolation of our current knowledge
to LHC energies and ask which physics questions can be addressed at the LHC.
\end{abstract}
\section{Introduction}

\begin{figure}[btp]
  \begin{center}
    \includegraphics[width=9cm]{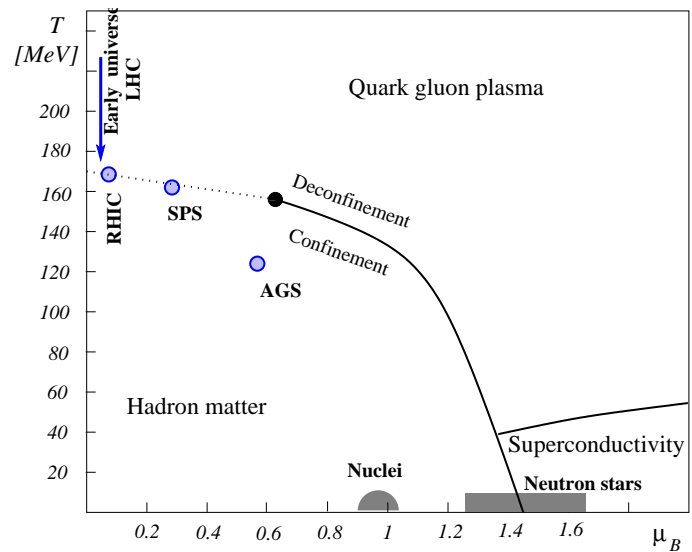}
    \caption{Schematic QCD phase diagram in the $T -\mu_B$ plane. 
    At low $T$ and $\mu_B$ nuclear matter shows confinement and hadrons
    are the degrees of freedom. At higher $T$ a phase 
    transition to a deconfined quark gluon plasma with restored chiral symmetry
    is predicted by lattice QCD. 
    The phase transition might exhibit a critical point at about $\mu_B \sim
    700$ MeV. More exotic quark phases can occur at high density, e.g.\ in the
    interior of very dense neutron stars. 
    Chemical freeze-out conditions reached in heavy ion 
    experiments at AGS, SPS and RHIC are also indicated. The blue arrow along 
    the $T$ axis shows how the matter is supposed to evolve at LHC before 
    freeze-out, starting at very high temperature. The evolution of the early 
    universe a few microseconds after the big bang took a similar 
    path.}
    \label{fig:phasediag}
  \end{center}
\end{figure}

The ALICE detector at LHC will be the next generation facility for exploring
the features of hot nuclear matter. The goal is to learn more about the 
universe as it existed a few microseconds after the big bang. A very hot and
dense phase of partons was present at these times. Later, a phase transition
occurred during which hadronic matter as we know it today was formed.
In the laboratory we try to create conditions that are similar to those 
in the early universe by colliding ultrarelativistic heavy ions, e.g. at the
RHIC facility and --- in a few years from now --- at the LHC. The hope is to 
create a deconfined quark gluon plasma (QGP) for a few fm/$c$ and to 
study its properties. Fig.\ \ref{fig:phasediag} shows our knowledge about QCD 
in the plane of temperature $T$ and baryon chemical potential $\mu_B$ 
with points that were tested in heavy ion experiments so far.
Also the region covered by the LHC, which lies at small baryon chemical
potential but at temperatures well above the phase transition
temperature of $T_c\sim 170$ MeV \cite{Karsch:01} is shown.

The theoretical knowledge about the QCD phase transition so far is mainly 
coming from lattice QCD (see e.g.\ \cite{Karsch:2001cy} for a review). 
It is possible to simulate the behavior of QCD along 
the $T$-axis for $\mu_B=0$. It is found that the scaled energy density 
$\epsilon/T^4$ steeply increases at $T_c$.
Since
\begin{equation}
  \epsilon = g_{\text{DOF}}\frac{\pi^2}{30} T^4,
\end{equation}
this can be understood by the rapidly rising number of degrees of freedom 
$g_{\text{DOF}}$ when the confined partons are liberated.
Lattice QCD also shows that the deconfinement phase transition occurs together
with a restoration of the spontaneously broken chiral symmetry. Recently
calculations at finite $\mu_B$ became available \cite{Fodor:2001pe}. 
They predict a critical point on the phase transition line that is shown in 
Fig.\ \ref{fig:phasediag}.

\section{Signatures of the phase transition}

If a quark gluon plasma is created in a collision of two large nuclei, it
eventually has to hadronize again. The detectors in our experiments can only
measure the hadronic debris from this collision. A direct observation of the
plasma is not possible. There has been a long discussion over the past
25 years what the possible signatures of a phase transition from a quark 
gluon plasma might be. We give a short (and probably incomplete) list of some
of the more popular ideas:
\begin{itemize}
\item Indirect measurements of thermodynamical quantities like the latent
  heat of the phase transition.
\item The enhanced production of strange particles.
\item Modifications of hadron properties, like the mass and width of the
  $\rho$ meson, through the presence of hot nuclear matter.
\item The detection of thermal photon or dilepton radiation from a thermalized
  QGP.
\item The suppression of quarkonia like the $J/\Psi$.
\item The energy loss of fast partons in a QGP, the so called jet quenching.
\item Fluctuations in net charge or baryon number.
\item Collective vacuum excitations like the disoriented chiral condensate
  (DCC).
\end{itemize}
In the following we discuss some of these ideas.

\subsection{Latent heat}
One of the simplest approaches one can think of is to measure quantities 
that are thermodynamically related to the phase transition, like the latent 
heat. A very interesting observation has been made regarding the slope of 
hadron spectra. 
Hadron spectra at low transverse momenta --- below several GeV/$c$, can be 
described phenomenologically by a blast wave fit \cite{Schnedermann:1993ws}. 
This assumes a thermalized spectrum with an effective temperature $T$. 
The spectrum is additionally boosted by a collective radial flow. 
The data collected for kaons at different energies seem to indicate that a 
plateau in the temperature was reached at SPS energies, while at higher RHIC 
energies the temperature is again rising. This could be interpreted as latent 
heat that is used to free the additional partonic degrees of freedom.

\subsection{Strangeness enhancement}
One of the classical signatures of the QGP is strangeness enhancement 
\cite{Rafelski:pu}. Since
the strangeness content of the colliding nuclei in the initial state is nearly
negligible, all strange particles in the final state have to be generated
in the collision. In purely hadronic matter, e.g.\ kaons and 
$\Lambda$s have to be produced through hadronic channels. In a quark gluon 
plasma, $s\bar s$ pairs will be created. This is enhanced by the relatively 
small strange quark mass $m_s < T_{\text{QGP}}$ in the chirally restored phase.
The strange quarks will then hadronize into strange hadrons and lead to higher 
yields compared to the hadronic scenario. This can already be seen at the
CERN SPS \cite{Fanebust:xb}.

\subsection{$J/\Psi$ suppression}
It was argued by Matsui and Satz \cite{Matsui:1986dk}, that the disappearance
of quarkonia states is a signal for the QGP. Lattice calculations show that
the heavy quark potential is effectively screened in a plasma above $T_c$,
so that $c\bar c$ bound states are melting \cite{Kaczmarek:2002mc}. 
It is then unlikely that the $c\bar c$ will find together at hadronization. 
Instead, they will mostly end up in open charm states. However, already at 
RHIC energies charm quarks could be so abundant, that spontaneous 
recombination of $c\bar c$ pairs could occur if $c$ quarks thermalize in the
medium \cite{Rapp:2003wn}. That would
alter the suppression scenario and could even lead to a $J/\Psi$ enhancement.

\subsection{Jet quenching}
\begin{figure}[btp]
  \begin{center}
    \includegraphics[width=7cm]{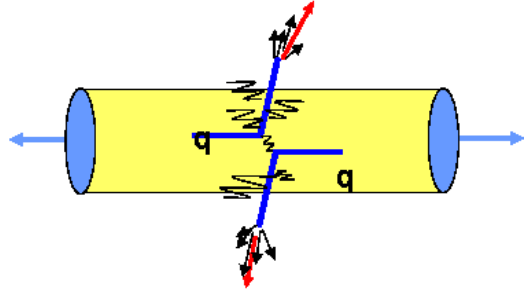}
    \caption{The jet quenching mechanism: Partons produced in hard QCD 
      processes suffer from final state interactions with the surrounding 
      hot medium. Induced bremsstrahlung leads to energy loss.}
    \label{fig:enloss}
  \end{center}
\end{figure}

\begin{figure}[btp]
  \begin{center}
    \includegraphics[width=8cm]{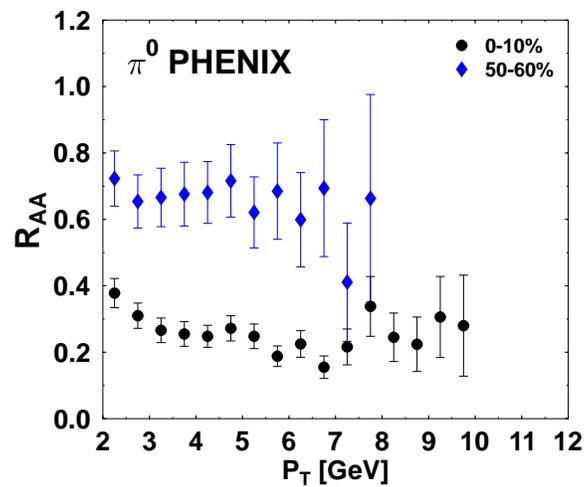}
    \caption{$R_{AA}$ for $\pi^0$ measured by the PHENIX collaboration 
      \cite{Adler:2003pi0} in Au+Au collisions at RHIC for two different 
      centrality bins. A large suppression can be observed for central 
      collisions}
    \label{fig:raa}
  \end{center}
\end{figure}

Fast partons produced in hard QCD interactions between the two colliding nuclei
have to travel through the hot and dense medium surrounding them. It was
proposed very early by Bjorken \cite{Bjorken:1982tu}, that these partons
lose energy by interactions with the medium, see Fig.\ \ref{fig:enloss}. 
This "jet quenching" was studied
in a series of theoretical papers over the last 10 years \cite{Thoma:1990fm}. 
The main process is the induced radiation of gluon bremsstrahlung by
interactions with the hot nuclear matter. This results in a suppression 
of hadron production at $P_T > 2$ GeV. Furthermore, since for strong energy
loss the emission of high-$P_T$ hadrons is dominated by surface emission, 
the correlated backside jet nearly vanishes. This was confirmed at RHIC 
\cite{PHENIX,Adler:2003pi0}. Fig.\ \ref{fig:raa} shows the 
nuclear suppression factor $R_{AA}$ for pions which is defined as the particle 
yield in Au+Au collisions normalized by the yield in $p+p$ collisions scaled 
by the number of binary nucleon-nucleon collisions. 
If a heavy ion collision were just a 
superposition of individual nucleon-nucleon collisions, $R_{AA}$ would be 1.
In recent $d+$Au control experiments, where no QGP is expected to be produced, 
indeed, no suppression in the hadron yield and back-to-back correlations 
similar to $p+p$ collisions were found \cite{Adler:2003ii}.

Theoretically, the effect of energy loss for hadron spectra can be described
by medium modifications in the fragmentation function of a parton $a$ into
hadron $h$ \cite{GW:00}. This modification approximately corresponds to a 
rescaling of the variable $z$
\begin{equation}
  D_{a\to h} (z,Q^2) \longrightarrow \tilde D_{a\to h} (z,Q^2)
  \approx D_{a\to h} \left(\frac{z}{1-\Delta E/E},Q^2\right).
\end{equation}
Theory predicts that the energy loss is quadratically dependent on the
length of the medium due to the so-called non-abelian 
Landau-Pomeranchuk-Migdal interference effect \cite{Landau:1953um}.

\section{Surprising RHIC results}

Besides the search for standard QGP signatures and the confirmation of
long standing theoretical expectations, experimental results from 
RHIC have provided many surprises. In this section we want to discuss the
anomalous baryon enhancement and the elliptic flow pattern found at RHIC.

It was widely assumed that hadron production at RHIC for $P_T > 2$ GeV can be 
described by perturbative QCD. This seemed to work well for pions if all known
nuclear corrections, like shadowing and energy loss, are taken into account.
On the other hand, protons and antiprotons deviate from this behavior. 
In a scenario with partonic energy loss, the suppression of $p$ and $\bar p$
should be the same as for pions. Furthermore, pQCD would predict a 
ratio $p/\pi^0 < 0.2$. RHIC results show that the $p/\pi^0$ ratio is 
about 1 between 1.5 and 4.5 GeV/$c$, and that there is nearly no nuclear
suppression in the yield of protons in this transverse momentum region 
\cite{Adler:2003kg}. 
A similar behavior was found in the strangeness sector for $\Lambda$s and 
kaons \cite{Adams:2003am}. 

An important quantity that can be measured in heavy ion collisions is the 
azimuthal anisotropy. At nonzero impact parameter $b>0$, the overlap zone of 
the two nuclei is not spherically symmetric. The initial anisotropy translates 
into an anisotropy of the final hadron spectra. This can be quantified by an 
expansion of the spectrum into harmonics
\begin{equation}
  \frac{d N}{2\pi P_T d P_T d\phi} =  \frac{d N}{2\pi P_T d P_T}
  \left( 1 + v_1(P_T) \cos \phi + 2 v_2(P_T) \cos 2\phi
  + \ldots \right) .
\end{equation}
The coefficient $v_2$ describes the elliptic anisotropy in the spectrum.
$v_2$ has been measured in RHIC experiments and shows a surprising dependence
on the hadron species. As a function of $P_T$ it rises and saturates above 
2 GeV/$c$. The value of saturation is always larger for baryons compared 
to mesons \cite{Adams:2003am}. 
It is believed that at these and higher values of $P_T$ the 
mechanism for translating the initial anisotropy into a final one is again 
partonic energy loss. Partons going into a direction where the interaction
zone is less extended will suffer less energy loss. However, this mechanism
should be blind to the hadron species.

The results above suggest that the range of validity for leading twist
perturbative QCD calculations only starts at higher transverse momentum.
Instead, the baryon enhancement and the azimuthal flow pattern can be 
understood in a simple picture of parton recombination \cite{DasHwa:77}. 
In pQCD hadronization happens through fragmentation. A single parton with 
momentum $p$ splits into gluons and $q\bar q$ pairs that eventually form 
hadrons. One of these hadrons is then measured and has momentum $P=zp$, 
($z<1$). The hadron spectrum is given by
\begin{equation}
  E\frac{dN_h}{d^3 P} = \sum_a
  \int\limits_0^1 \frac{dz}{z^2} \> D_{a\to h} (z) E_p \frac{dN_a}{d^3 p}
\end{equation}
where $dN_a/d^3 p$ is the spectrum of partons $a$ and $p= P/z$ is the parton
momentum.

On the other hand, if phase space is already densely populated with partons,
these can simply recombine to give mesons $M$ and baryons $B$ 
\begin{equation}
  q \bar q \rightarrow M, \quad qqq \rightarrow B, \quad \bar q\bar q\bar q
  \rightarrow \bar B.
\end{equation}
In this case, the momenta of the valence partons add up. 
Quantitatively this can be formulated in a coalescence formalism using hadron
wave functions $\phi_h$ in light cone coordinates \cite{Fries:2003kq}. For
a meson $M$ the spectrum can be written as
\begin{equation}
  E\frac{dN_M}{d^3 P} = C_M \int\limits_\Sigma d\sigma \frac{P\cdot u(\sigma)}{
  (2\pi)^3} \int\limits_0^1 dx \> w_q (\sigma; xP^+) \left| \phi_M(x) \right|^2
  w_{\bar q} (\sigma; (1-x)P^+)
\end{equation}
where $C_M$ is a degeneracy factor, $w_q$ and $w_{\bar q}$ are the phase 
space distributions of the recombining quark and antiquark respectively,
$x$ is the momentum fraction of the quark in light cone coordinates and
$\Sigma$ is the hadronization hypersurface in Minkowski space.

It can be shown that recombination is always more effective than 
fragmentation for an exponential parton spectrum, but fragmentation will 
dominate at high $P_T$ for a parton spectrum in power law form.
Furthermore, for an exponential parton spectrum $w= e^{-P^+ /T}$,
we have 
\begin{equation}
  w_q (xP^+) w_{\bar q} ((1-x)P^+) = e^{-P^+/T}
\end{equation}
in the case of a meson $M$ and analogously 
\begin{equation}
  w_q (x_1 P^+) w_{q} (x_2 P^+) w_{q} ((1-x_1-x_2) P^+) 
  = e^{-P^+/T}
\end{equation}
in the case of a baryon $B$. $P^+$ here always denotes the large
component of the hadron momentum on the light cone. Hence recombination
naturally provides a ratio $B/M \sim 1$. We should add that the description
of recombination given above is only valid if $P^+$ is much larger than 
$\Lambda_{\text{QCD}}$ and the masses of the hadrons.

\begin{figure}[btp]
  \begin{center}
    \includegraphics[width=8cm]{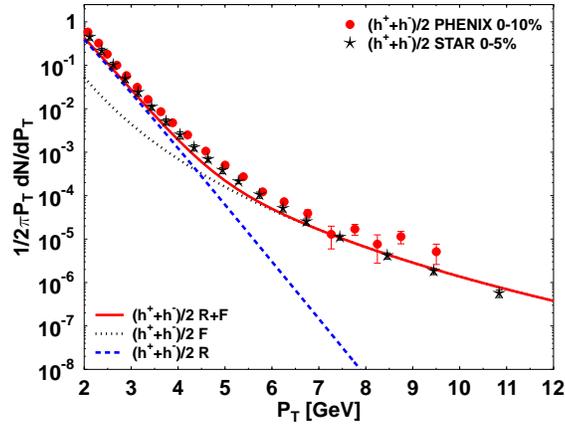}
    \caption{The spectrum of charged hadrons $(h^+ + h^-)/2$ at midrapidity
    for central Au+Au collisions at RHIC. The result of a leading order pQCD
    calculation with energy loss (dotted line), recombination of a thermalized
    parton phase (dashed line) and the sum of both (solid line) are shown.
    Data are from STAR \cite{Adams:2003kv} and PHENIX 
    \cite{Velkovska:2003sqm}.}
    \label{fig:chhadrhic}
  \end{center}
\end{figure}

Fig.\ \ref{fig:chhadrhic} shows the result of a calculation for charged hadrons
at RHIC using recombination from a thermalized phase of constituent quarks
with a temperature $T=175$ MeV and an average radial flow velocity 
$v_T = 0.55c$ and fragmentation in a leading order pQCD calculation including
energy loss. One can clearly see the two different domains of hadron 
production. Only above $P_T = 5$ GeV/$c$ is leading twist pQCD a good 
description for hadron production. Below that recombination is important.
Data on the $p/\pi^0$ ratio in this momentum range are well reproduced by this 
calculation.

For the recombination of partons, one can derive a simple scaling law that
connects the azimuthal anisotropy $v_2$ for partons $p$ and hadrons $h$
\cite{Fries:2003kq,LinKo:02}
\begin{equation}
  v_2^h (P_T) = n v_2^p (P_T /n)
\end{equation}
where $n=2$ for mesons and $n=3$ for baryons. This scaling law which was 
recently confirmed by the STAR and PHENIX collaborations \cite{Adler:2003kt},
can be considered as a new `` smoking gun'' for the creation of a quark gluon 
plasma.
The scaling law visualizes the anisotropic flow in the parton phase before
hadronization.

\section{What's different at LHC?}

At LHC lead nuclei will collide with a center of mass energy 
$\sqrt{s}=5.5$ TeV.
This leads to a much higher initial energy density. Higher energy density
and increased lifetime will make initial state effects less important, but 
will also enhance the role of the QGP phase over final state hadronic 
interactions. On the other hand, probes with very high transverse momentum,
up to 100 GeV/$c$, will be available and it might be possible to measure
jets \cite{cernyr:2003}. Also heavy $c$ and $b$ quarks are plentiful probes 
at this energy.

\subsection{Saturation physics}
The relevance of saturation of the nuclear wave function \cite{Gribov:1981ac} 
has already been discussed at lower energies. It is clear that LHC will be
the ideal testing ground for saturation physics. The basic idea is that
the gluon distribution in a QCD bound state cannot continuously grow fast
at small
Bjorken $x$ without violating unitarity. At some point, gluon fusion will
balance the growth. The scale at which the probability of gluon interactions 
in the nucleus wave function becomes of the order of one determines the 
saturation scale $Q_s$ \cite{Gribov:1981ac}
\begin{equation}
  \frac{x G_A(x,Q_s^2)}{\pi R_A^2} \frac{\alpha_s(Q_s^2)}{Q_s^2} \sim 1.
\end{equation}
From this equation one obtains the scaling with the nuclear size 
$Q_s^2 \sim A^{1/3} x^{-0.5}$. 
Starting from these assumptions various phenomenological consequences have 
been derived \cite{McLerran:1993ni} for QCD at large $\sqrt{s}$. See e.g.\ 
\cite{Iancu:2002xk} for an overview.

Saturation has also to be taken into account, when one of the fundamental 
questions is addressed: What are the particle multiplicities we can expect 
at LHC? This is very important for the detector design.
Estimates have been given in \cite{Eskola:1999fc} using perturbative QCD
and saturation.

\subsection{Jet quenching at the LHC}
It is expected that the effect of jet quenching at LHC is even larger than
at RHIC due to the higher energy density in the medium. Estimates depending
on the produced particle multiplicity show that quenching factors of
10--30 can occur for pions at $P_T=10$ GeV \cite{Vitev:2002pf}. Nevertheless,
hard QCD will be an important part of the heavy ion program at LHC. It
might be possible to measure jets directly, as it is done in $p+\bar p$ and 
$e^+ +e^-$ collisions. That eliminates the uncertainty coming from 
fragmentation functions in theoretical calculations and could be a valuable 
contribution to our understanding of energy loss in the medium 
\cite{cernyr:2003}.

\subsection{Recombination at the LHC}
\begin{figure}[tbp]
  \begin{center}
    \includegraphics[width=8cm]{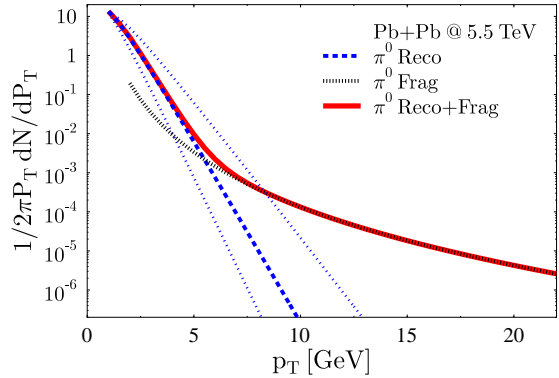}
    \caption{Transverse momentum spectra of $\pi^0$
    for central Pb+Pb collisions with $\sqrt{s}=5.5$ GeV at midrapidity. 
    Fragmentation from pQCD (dotted), recombination (long dashed) and
    the sum of both (solid line) are shown.
    The parameters for the thermal parton phase are $T=175$ MeV and $v_T = 
    0.75c$. For pions recombination for different radial flow velocities
    0.65$c$ and 0.85$c$ (short dashed, from below) are also shown.}
    \label{fig:lhcspec}
  \end{center}
\end{figure}

\begin{figure}[tbp]
  \begin{center}
    \includegraphics[width=8cm]{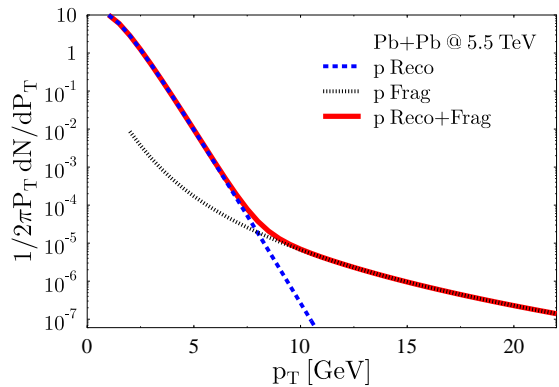}
    \caption{The same as Fig.\ \ref{fig:lhcspec} for the transverse momentum
      spectrum of protons ($v_T=0.75 c$ only).}
    \label{fig:lhcspecprot}
  \end{center}
\end{figure}

\begin{figure}[tbp]
  \begin{center}
    \includegraphics[width=8cm]{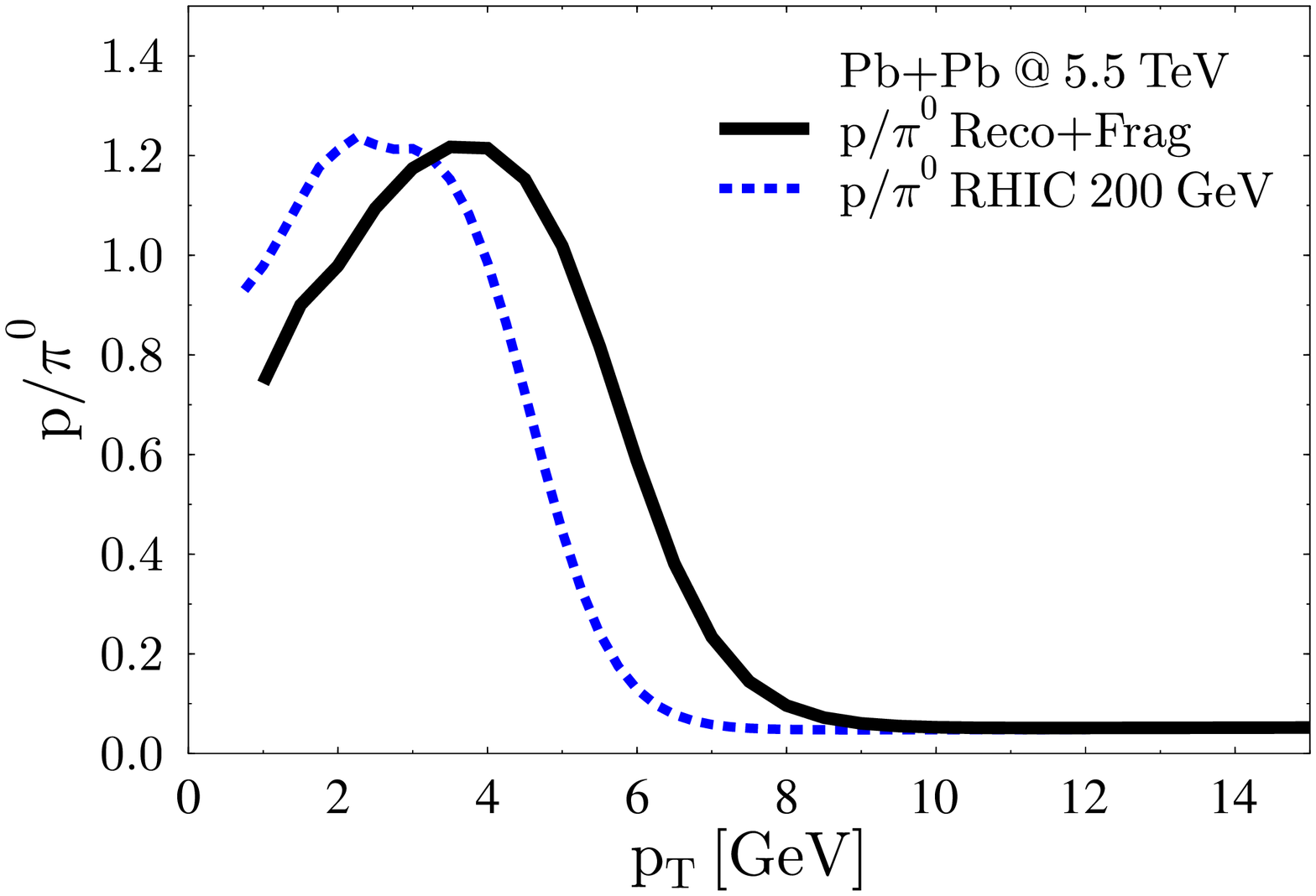}
    \caption{The $p/\pi^0$ ratio for Pb+Pb at LHC (solid) and for Au+Au at
    RHIC (dashed line) as predicted by a calculation using recombination
    and pQCD. The baryon enhancement is pushed to higher $P_T$ for LHC.}
    \label{fig:lhcratio}
  \end{center}
\end{figure}

The large nuclear suppression factor raises the question whether ``soft'' 
thermal physics will push its limits to even higher $P_T$ at LHC. Preliminary
studies with recombination from a thermalized parton phase confirm this.
Figs.\ \ref{fig:lhcspec},\ref{fig:lhcspecprot} shows an estimate of $\pi^0$ 
and $p$ spectra at LHC taking into account recombination and fragmentation  
\cite{Fries:2003kq}. The energy loss used is in accordance with the estimates
of Gyulassy and Vitev \cite{Vitev:2002pf} and for the thermal parton phase
a temperature $T=175$ MeV and radial flow $v_T = 0.75c$ are assumed.
The crossover between the recombination domain and the pQCD domain is shifted
to 6 GeV/$c$ (from 4 GeV/$c$ at RHIC \cite{Fries:2003kq}) for pions
and to 8 GeV/$c$ (from 6 GeV/$c$ at RHIC) for protons.
Correspondingly the $p/\pi^0$ ratio at LHC is also shifted to higher $P_T$,
see Fig.\ \ref{fig:lhcratio}.

\subsection{Measurements of medium properties}
\begin{figure}[tbp]
  \begin{center}
    \includegraphics[width=7cm]{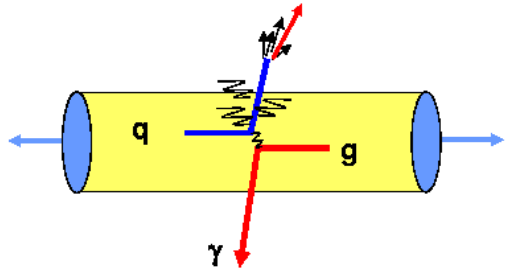}
    \caption{Tagging of a jet or leading hadron by a photon on the other
    side: The photon can escape without final state interactions and carries 
    information about the transverse momentum of the simultaneously
    produced parton before that loses energy in the medium.}
    \label{fig:jetphottag}
  \end{center}
\end{figure}

Once partonic energy loss is established, one would like to measure 
properties of the QGP by using hard QCD probes. One sort of precision 
measurements could be photon-tagged jets or hadrons. 
For this one considers the back-to-back production of a parton and a photon,
e.g.\ $q+g\to q+\gamma$ as shown in Fig.\ \ref{fig:jetphottag}. 
Due to the weakness of electromagnetic interactions,
the photon can leave the medium unaffected, while the outgoing parton will
suffer from final state interactions. By momentum conservation the
transverse momentum of the photon is equal to the initial momentum of the 
outgoing parton directly after the production. But the parton will suffer from
energy loss before it can hadronize. 
By tagging a reference photon and looking at hadrons or jets at the
opposite side, precision measurements of energy loss will be possible
\cite{Wang:1996yh}. The same measurement can be done with back-to-back 
production of a parton and a virtual photon, which then decays into a lepton 
pair \cite{Srivastava:2002kg}.

\begin{figure}[tbp]
  \begin{center}
    \includegraphics[width=7cm]{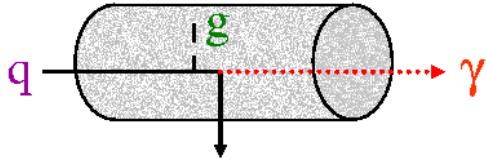}
    \caption{Jet photon conversion: A fast parton is turned into a photon 
      with comparable momentum by interactions with the medium.}
    \label{fig:jetphotconv}
  \end{center}
\end{figure}

\begin{figure}[tbp]
  \begin{center}
    \includegraphics[width=9cm]{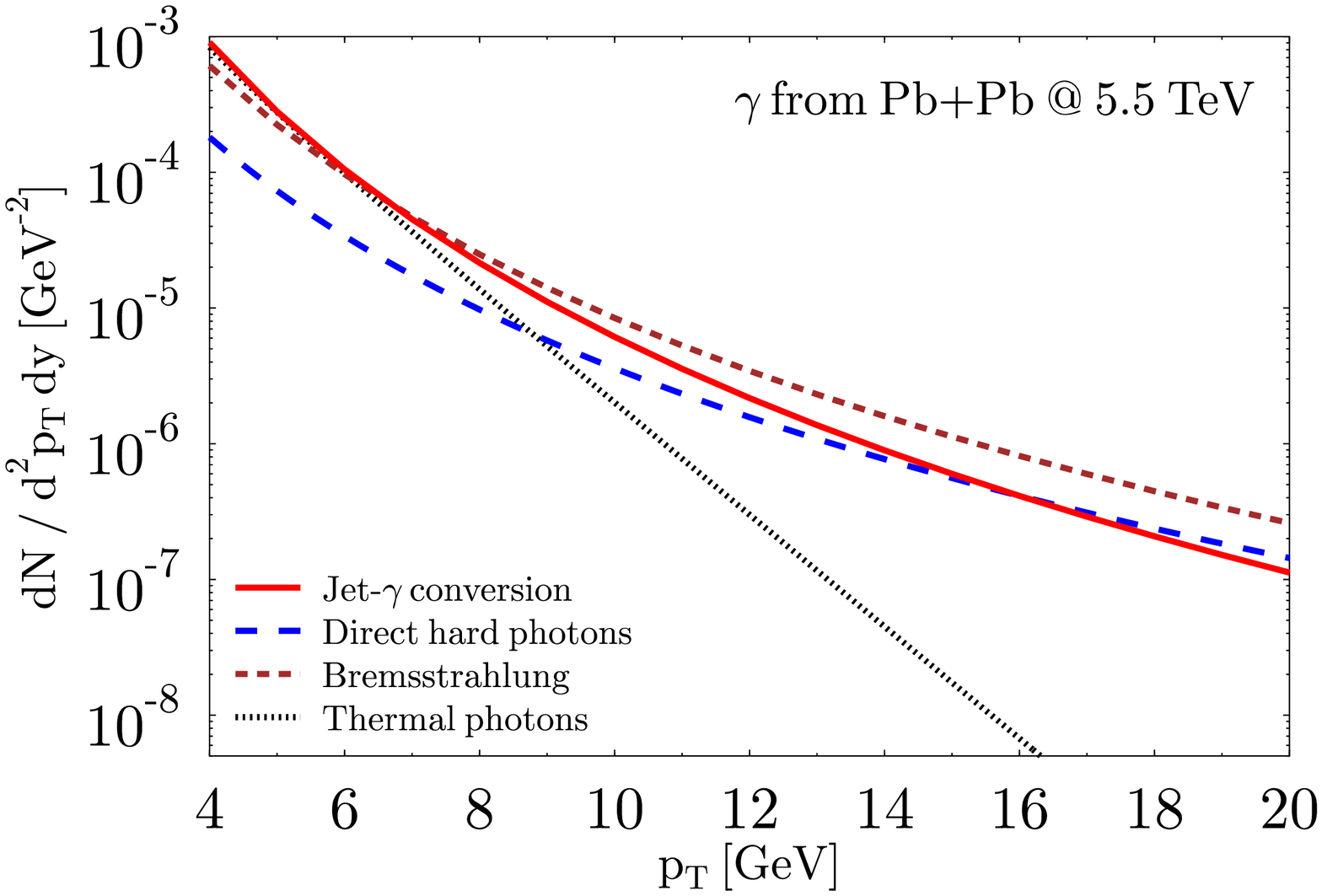}
    \caption{Photons from jet-photon conversion (solid line) at LHC compared 
     to other direct photon sources: thermal radiation from the plasma 
     (dotted), direct photons from primary hard interactions (long dashed) and 
     bremsstrahlung from primary hard interactions (short dashed line).
     The yield from conversion is of the same order of magnitude as the other
     processes. See \cite{Fries:2002kt} for further details.}
    \label{fig:convlhc}
  \end{center}
\end{figure}

Another promising idea is to measure induced photon radiation from fast 
partons interacting with the medium \cite{Fries:2002kt}. It is  
characteristic for the Born cross sections for these processes, e.g.\ 
\begin{equation}
q(\text{jet}) + \bar q(\text{medium}) \to g + \gamma \> \text{or} \> 
\quad  q(\text{jet}) + g (\text{medium}) \to q + \gamma \, ,
\end{equation}
--- see Fig.\ \ref{fig:jetphotconv} --- that they peak in forward and backward
directions (in the center of mass frame). That means that the three-momentum of
the photon is very close to one of the two initial momenta. If one restricts 
the measurement of the photon to large transverse momentum, 
say $P_T > 4$ GeV/$c$, then
the momentum of the photon will be that of the fast parton. This jet-photon
conversion mechanism is another opportunity to measure the momentum of 
a fast parton in the plasma. This time the photon emission will take place
while this parton is traveling through the plasma and, in order to obtain the 
photon rate, one has to integrate over the path of the parton. Therefore this
measurement is sensitive 
to the evolution of density and temperature of the medium. It has been 
estimated that photons from this conversion mechanism are shining quite 
brightly 
at RHIC and LHC compared to other photon sources, cf.\ Fig.\ 
\ref{fig:convlhc}. Again, similar measurements are possible with dileptons 
instead of photons \cite{Srivastava:2002ic}.

\section{Summary}

We have discussed several key issues of heavy ion physics. While we might
already discover the quark gluon plasma at RHIC, it will be at LHC that we
can systematically study its properties by making use of the the plentiful
hard probes. Some important questions have to be addressed. 
\begin{itemize}
\item How does gluon saturation work at small $x$?
\item How exactly are fragmentation functions modified in the medium?
\item How does the energy loss depend on the energy density?
\item Are heavy quarks thermalized?
\item What is the role of nuclear higher twist effects?
\end{itemize}
The LHC will be the ideal machine to provide answers.

\end{document}